\documentclass[floatfix,amssymb,amsmath,
a4paper,reprint,11pt,onecolumn,tightenlines]{revtex4-2} 

\usepackage{amsmath}
\usepackage{listings}
\usepackage[colorlinks]{hyperref}
\usepackage{color}
\usepackage{graphicx}

\usepackage{latexsym} 
\usepackage{amssymb} 

\begin{document}
\title{Dynamics of a bricklayer model: multi-walker realizations of true
  self-avoiding motion}

\author{A.~C.~Maggs} \affiliation{CNRS UMR 7083, ESPCI Paris, Universit\'e PSL,
  10 rue Vauquelin, 75005 Paris, France.}

\begin{abstract}
  We consider a multi-walker generalization of the true self-avoiding walk: the
  bricklayer model. We perform stochastic simulations, and solve the 
  partial differential equations that describe the collective motion of $N$
  bricklayers/walkers coupled to the contour of an expanding wall.
  In the large-$N$ limit, the results from simulation agree
  with the solution of the partial differential equations.
\end{abstract}
\maketitle
\subsection*{Introduction}
The true self-avoiding walk~\cite{Amit} was introduced to describe a growing
polymer on a lattice, where the monomer to be added to the end of the walk tries
to avoid previously visited sites.  It was studied using various approximate
analytic tools, and it was concluded that the statistics of this walk are
distinct from those of an equilibrium polymer.  This self-avoiding process has
found applications in several fields, including chemotaxis~\cite{biophy,prx}, but
also as a tool to describe the large-scale dynamics of a family of
non-reversible Monte Carlo algorithms~\cite{Wilson, Kapfer, Manon, LeiFF,
  nonrev, Maggs2025}. There are also direct links to a lifted ``totally
symmetric simple exclusion process'' (TASEP)~\cite{essler2023lifted, brune}.

Great progress in understanding this process was made in a series of
papers~\cite{Toth, Wendelin, DUMAZ20131454, 2025convergence}, which give the full
analytic solution to the motion in one spatial dimension. In particular, it was
demonstrated that the motion explores a spatial extent growing in time as
$t^{2/3}$. These papers give explicit results for the distribution function of
the end-to-end separation of the walk.

The founding paper~\cite{Amit} of this model also studied the process in the
continuum with the coupled equations
\begin{eqnarray}
  \frac{d {X(t)} } {d t } &=& -\frac{\partial  h( {X}(t), t)} {\partial x}\, ,  \label{eq:motion} \\
  \frac {dh({x},t ) }{dt} &=&\,\,  \delta ( {x} - {X(t)})\, .  \label{eq:density}
\end{eqnarray}
In eq.~(\ref{eq:motion}) the function $h(x,t)$ measures the number of previous
visits to the position $X$, which acts as a repulsive potential in the motion.
In eq.~(\ref{eq:density}) the repulsive potential is itself modified by 
the walker's visits to the position $X(t)$. We note that we have changed notation
compared to certain papers, including our own, to remain consistent
with~\cite{Toth2002}, the main inspiration of the present paper.

Eq.~(\ref{eq:motion}) describes the motion of a particle moving at velocity
$u=-{\partial h(t, {X}(t))}/ {\partial x} $, so it corresponds to the continuity
equation
\begin{equation}
  \frac{\partial \rho} {\partial t} + \frac{\partial (u  \rho)}{\partial x}=0
  \, , \label{eq:cont1}
\end{equation}
where $\rho=\delta(x-X(t))$ is the particle density.  If we now consider the
spatial derivative of eq.~(\ref{eq:density}) we find
\begin{equation}
  \frac{\partial u} {\partial t} + \frac{\partial  \rho}{\partial x} =0 \, . \label{eq:cont2}
\end{equation}
The properties of the coupled equations (\ref{eq:cont1},~\ref{eq:cont2}) for a
collection of many walkers are the main subject of the present paper. Many
properties of these equations are established by~\cite{Toth2002}, where
these equations are shown to be the continuum limit of a discrete model, which
the authors interpret as a company of bricklayers building a long wall. 
To build a uniform wall, masons preferentially move down the local gradient of
height $h$, leaving one brick on the wall each time they move. One is led to
consider the collective hydrodynamic behavior of $N$ independent bricklayers
described by a continuum density distribution, $\rho(x)$, as a natural
generalization of the true self-avoiding walk. The work of~\cite{Toth2002} is to
be placed in the context of other nonlinear models of interface growth, including
the model of Kardar, Parisi and Zhang (KPZ)~\cite{kpz}, which also describes the
dynamics of an interface $h(x,t)$.

Below, we give the exact definition of the discrete model of
bricklayers and perform simulations on this model with a variable number of
bricklayers, $N$. We consider the analytic solution of the partial differential
equations for the case of bricklayers all starting together. We also study the
dynamics of bricklayers starting at random positions on a periodic, circular wall.

\subsection*{Bricklayers}
Let us recall, briefly, the discrete model of~\cite{Toth2002}. A wall is being
built with bricks added on the links $j,j+1$ of a one-dimensional lattice. The
number of bricks on this link is denoted $h_j$. Inspired by the continuum
equations, we also introduce the discrete negative gradient of $h_j$ such that
$z_j=h_{j-1} - h_j$. The bricklayers are on the nodes of the lattice with no
constraints on their occupation number. The number of bricklayers at site $j$ is
$n_j$.  At any moment, a bricklayer can jump to the right or left, leaving one
brick in the appropriate column. The jumps occur at a rate $r(z_j)$. One imposes
\begin{equation}
  \label{rcond}
  r(z)r(-z+1)=1, 
\end{equation}
so that the local jump rate depends on the local gradient of the wall's height,
corresponding to the number of previous visits of bricklayers.  In our numerical
work, we choose $r\sim\exp(\beta z)$ with $\beta =0.4$. When a bricklayer jumps,
the following changes of configuration may occur:
\begin{equation*}
  (n_j,z_j),(n_{j+1},z_{j+1})
  \to
  (n_j-1,z_j-1),(n_{j+1}+1,z_{j+1}+1)
\end{equation*}
with rate $n_j r(z_j)$, and
\begin{equation*}
  (n_j,z_j),(n_{j-1},z_{j-1})
  \to
  (n_j-1,z_j+1),(n_{j-1}+1,z_{j-1}-1)
\end{equation*}
with rate $n_j r(-z_j)$.

Both $\sum_j n_j=N$ and $\sum_j z_j$ are conserved, as is the
parity of $(n_j+z_j)$ for each site.  In the long-time limit, we expect to
generate smooth distribution functions. It is natural to identify the discrete
label $j$ and the continuum variable $x$. The discrete occupation number $n_j$
maps to the distribution $\rho(x)$ and $z_j$ is linked to $u(x)$.  We generate a
starting configuration and store the randomly generated event times in a
red-black tree using the C\texttt{++} \texttt{multiset} container. We find the
time of the first event, update the configuration, and recalculate times for sites that have
been modified. The process is repeated to evolve the system a total time
$t$.  

\subsection*{Scaling with number of bricklayers}
We consider a simple generalization of previously given scaling
arguments~\cite{Pietronero,Ottinger_1985} for the width of the function $\rho$ after time
$t$, starting with all builders on a single site. We consider a system with $N$
builders, with a characteristic spatial scale which is given by 
$\ell \sim N^\beta t^\alpha$ at time $t$.  The number of events is comparable
to $Nt$, which builds a wall of height $Nt/\ell$.  The gradient of height must
then scale as ${ \partial h}/{\partial x} \sim N t/\ell^2$. After acting for a
time $t$ on the bricklayers, this must give motion again comparable to
$N^\beta t^\alpha = Nt^2/(N^{2\beta}t^{2\alpha})$. This implies $\alpha=2/3$,
$\beta=1/3$, so that the extent of the distribution varies as
\[\ell \sim N^{1/3} t^{2/3}\, .\] The mean density in the occupied region then
varies as ${(N/t)}^{2/3}$. We expect these results to hold when $t\gg N$ so that
the mean occupation of bricklayers in the propagating region is smaller than
unity. We estimate that the time to explore a system of length $L$ scales
as $t \sim L^{3/2}/N^{1/2}$.

\subsection*{Simulation results}
We performed simulations to study the evolution of the functions $h(x,t)$ and
$\rho(x,t)$. We place all the bricklayers on a single starting site (the origin
$x=0$ in our figures) and at the end of the simulation record the empirical
distributions.  To generate high-quality data, we typically average over $10^5$
realizations of the process.  The data for different times and numbers of bricklayers 
display disparate scales; when we compare the distribution functions, we scale
the functions to unit area and unit variance, and denote the result
$\bar \rho(\bar x,t)$ and $\bar h(\bar x,t)$, with scaled coordinate $\bar x$
and scaled distributions $\bar \rho$ and $\bar h$.

\begin{figure}
  \includegraphics[width=.60\columnwidth]{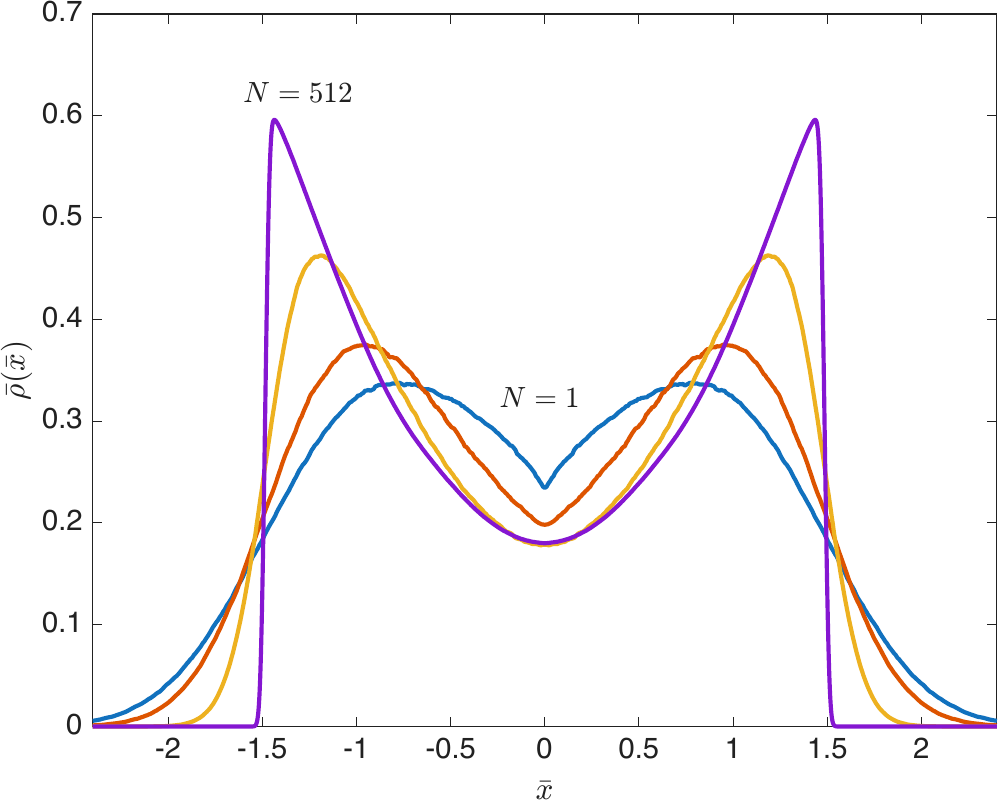}
  \caption{Evolution of the density profile with builder number, for fixed time
    $t=512$ and $N= 1, 4, 16, 512$ bricklayers. The case $N=1$ reproduces the
    known distribution of the true self-avoiding walk. Curves scaled to unit
    area and unit variance for comparison.} 
  \label{fig:final}
\end{figure}

\begin{figure}
  \includegraphics[width=.60\columnwidth]{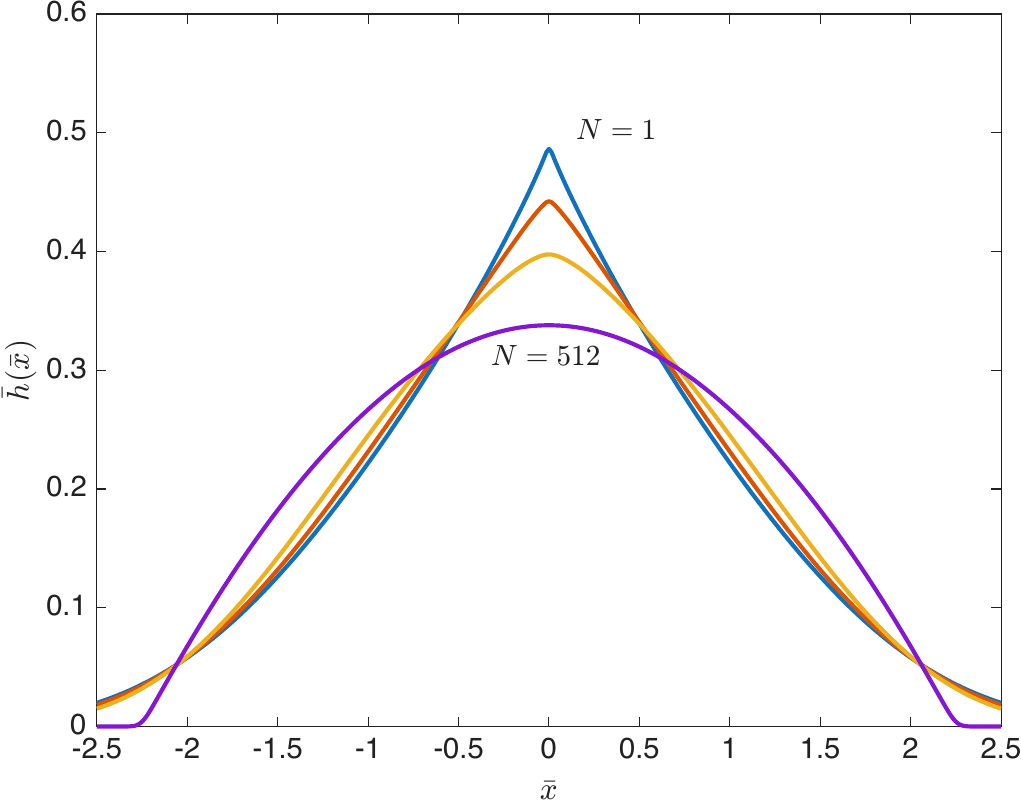}
  \caption{Evolution of the interface $\bar h(\bar x)$ with builder number $t=512$,
    $N= 1, 4, 16, 512$ builders. For large $N$ the distribution has a parabolic
    form, corresponding to the linear curve for $g(y)$ in
    Fig.~\ref{fig:compare}.  Curves scaled to unit area and unit variance for
    comparison. } 
  \label{fig:interface}
\end{figure}

We studied the case $N=1$ for which explicit results are
known analytically~\cite{DUMAZ20131454}, confirming that the function
$\bar \rho(\bar x,t)$ is correctly reproduced by the simulations. See  $N=1$
in Fig.~\ref{fig:final}. We then varied the number of bricklayers while keeping the
simulation time constant at $t=512$. The double-peak structure
familiar from $N=1$ is maintained, but the peaks are pushed to larger
separations;  the distribution of density  drops to zero more
abruptly for large values of $N$. The evolution of $h$ is shown in
Fig.~\ref{fig:interface}: For small $N$, the interface has a sharp maximum at
the origin (the site where the builders were placed), but this is replaced by a
broad maximum for $N$ large.

\begin{figure}
  \includegraphics[width=.60\columnwidth]{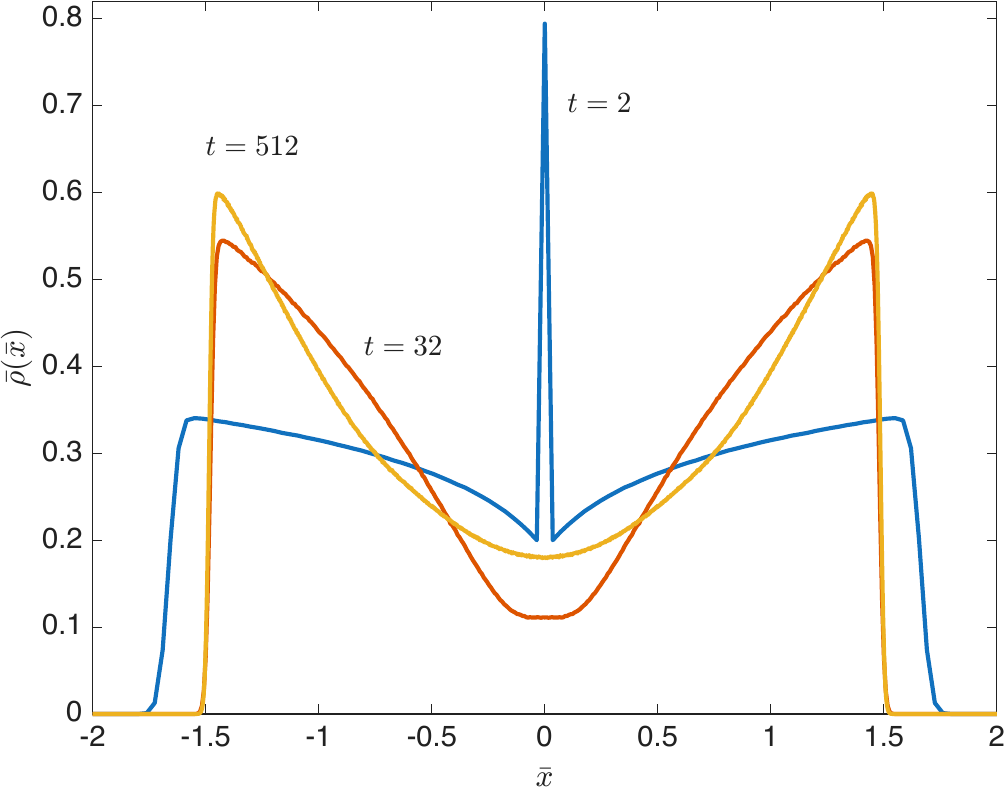}
  \caption{Evolution of the density $\rho$ with time for 512 builders, for
    times $t=2$, $t=32$, $t=512$.  For the shortest time (blue) a peak of
    density remains at the origin. This peak rapidly disappears (red), leaving a
    central depression in the density distribution. At the longest time (yellow)
    the density approaches an asymptotic parabolic form, with a rapid fall to
    zero at finite $\bar x$.
  } 
  \label{fig:interface2}
\end{figure}

We also studied the evolution in time when starting with $N=512$ bricklayers,
Fig.~\ref{fig:interface2}. At short times, $t=2$ there is a peak at the origin,
corresponding to builders that have not yet had time to move. By time $t=32$ a
dip has formed in the distribution function. The dip is shallower for the
longest time shown, $t=512$.  From our simulations, we were able to confirm the
law in $N^{1/3} t^{2/3}$ for the width of the distribution $\rho(x,t)$.

\subsection*{Scaling solution of partial differential equations}
Scaling analysis tells us that the coupled partial differential equations have a
solution of the form
\begin{align*}
  \rho = \frac{1}{t^{2/3}} f( x/t^{2/3}) \, ,\\
  u = \frac{1}{t^{1/3}} g( x/t^{2/3}) \, .
\end{align*}
These scaling relations are similar to those displayed by the solution to the
KPZ equation.  
Substituting into the
partial differential equations, we find coupled equations of $f$ and for $g$
which we write in matrix form:
\begin{equation}
  \begin{pmatrix}
    3 g -2y &  3 f\\
    3 & -2y
  \end{pmatrix}
  \begin{pmatrix}
    f' \\ g'
  \end{pmatrix}
  = \begin{pmatrix}
    2 f\\
    g
  \end{pmatrix} \, , \label{eq:coupled}
\end{equation}
where $y=x/t^{2/3}$. Clearly, there is a trivial solution to these equations:
$f(y)=g(y)=0$.
  
\begin{figure}
  \includegraphics[width=.60\columnwidth]{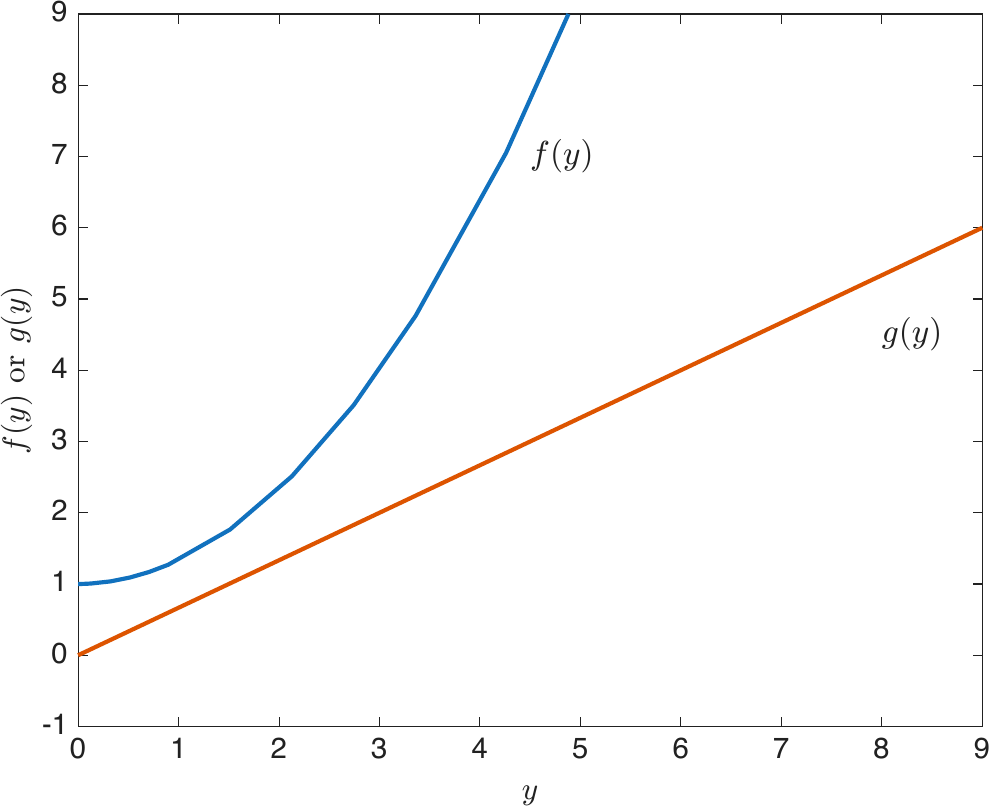}
  \caption{Integration of the coupled equations eq.~(\ref{eq:coupled}), starting
    from $f(0)=1$ and $g(0)=0$. The functions $f(y)$ and $g(y)$ do not decrease to
    zero for large $y$.}
  \label{fig:compare} 
\end{figure}

We used a Runge-Kutta integrator to solve eq.~(\ref{eq:coupled}) from initial values
of $f$ and $g$ at $y=0$. The value of $f(0)$ can
be chosen to be unity by a rescaling of $x$ and $t$, and the natural choice
for a non-singular odd function $g$ is $g(0)=0$. In Fig.~\ref{fig:compare} we
show the numerical solution of the coupled equations. We
find that to high accuracy $g(y)$ is a linear function of $y$, while $f(y)$ is
quadratic. These results are surprising, since for large $y$ both functions must
be zero; they describe the finite time evolution of a compact distribution of
bricklayers. We show below that this is possible if we introduce discontinuities in
the two functions, so that the solution jumps to the trivial solution
$f(y)=g(y)=0$.


\subsection*{Analytic solution of the differential equations}
The theory of hyperbolic partial differential equations is complicated
by the non-uniqueness of solutions in the presence of shocks or
discontinuities. The theory of the selection of the correct physical solution in
such cases is subtle and will not be attempted here. We rather construct a
solution from observation of the simulation data Fig.~\ref{fig:interface2},
informed by the numerical integration of the coupled equations
Fig.~\ref{fig:compare}.  Empirically, we understand that the long-time limit of
the function $\rho$ is parabolic as found by the numerical integration, but it
is truncated to zero at some finite $y=y_0$, as shown by the simulations.
 
From the numerical integration for the scaling functions, we see that
$g= \gamma y$. If we choose $3 \gamma =2$ the equations are simple, and we only
need to solve
\begin{equation}
  3f' = 2y g' + g \label{eq:singular}
\end{equation}
for $f$ giving
\begin{math}
  f = A+y^2/3
\end{math}.  Let a discontinuity occur at $y=y_0$, then balancing the
singularities in eq.~(\ref{eq:singular}) gives $3\Delta f-2y_0 \Delta g=0$,
where $\Delta f$ and $\Delta g$ are the jumps in the respective functions. This
allows us to relate $A$ and $y_0$ giving
\begin{math}
  A= y_0^2/9
\end{math}.  We also have that the total number of bricklayers is given by
\begin{equation}
  \int ^{y_0}_{-y_0}  f(y) dy  =   2 y_0 A + 2 y_0^3/9 = N
\end{equation}
so
\begin{equation}
  y_0 = {\Big ( \frac{9 N}{4} \Big )}^{1/3}\, ,
\end{equation}
agreeing with the previous scaling argument for the width of the distribution as
a function of $N$. 
We integrate $g(y)$ to find the wall profile: we have
$-\partial h/\partial x = (1/t^{1/3})(2/3) (x/t^{2/3}) = 2x/(3t)$ so that
\begin{math}
  h(x,t)= B -\frac{1}{3} x^2/t
\end{math}
with the zero of $h$ at $x= t^{2/3} y_0$. Thus, $B= (1/3) t^{1/3} y_0^2$. 
The scaling form of the solution to the partial differential equations is
\begin{align}
  h(x,t)=& \frac{t^{1/3}} {3} \left ( y_0^2 -  {\left (\frac{x}{t^{2/3}}\right )}^2 \right )
           \Theta(y_0 -|x/t^{3/2}| )\, ,\\
  \rho(x,t) =& \frac{1}{9 t^{2/3}} \left  (y_0^2+ 3 {\left ( \frac{x}{t^{2/3}}
               \right)}^2 \right )
               \Theta(y_0 -|x/t^{3/2}| ) \, ,\\
u(x,t) = & \frac{2 x}{3t}  \Theta(y_0 -|x/t^{3/2}| ) \, .
               \label{eq:dist_rho}
\end{align}
The equations display discontinuous behavior at $(x/t^{2/3})=y_0$.

\begin{figure}
  \includegraphics[width=.60\columnwidth]{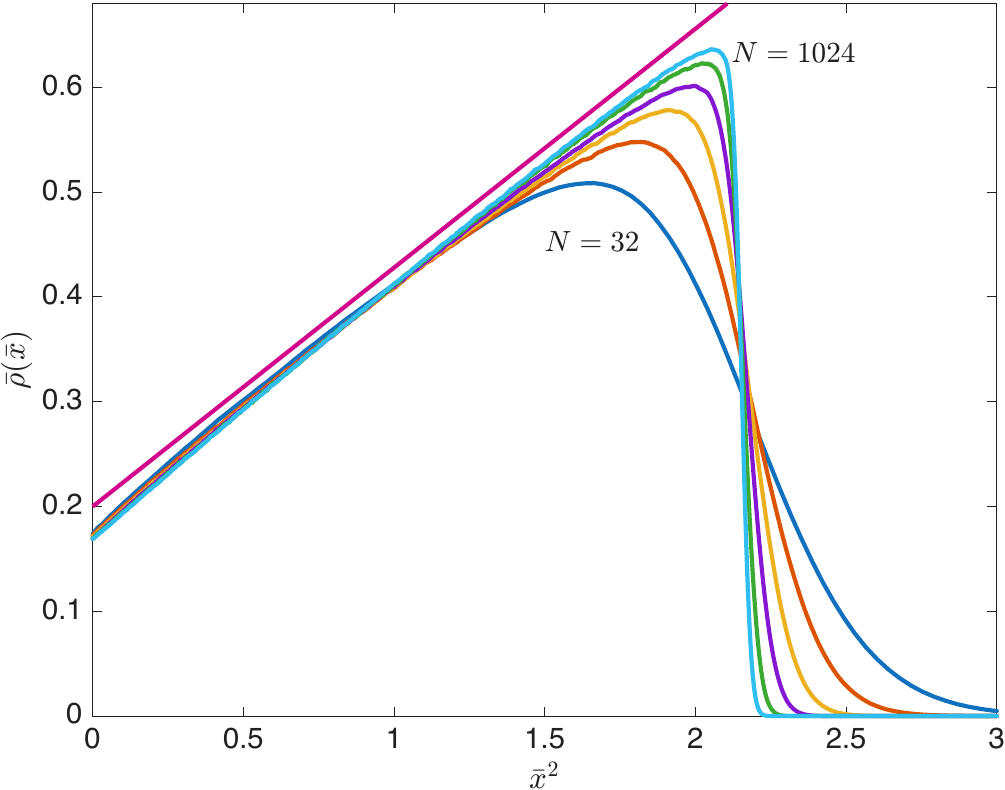}
  \caption{Density distributions for $N$=32, 64, 128, 256, 512, 1024. Plotted on a
    quadratic scale to show convergence of the distributions to triangular form
    as $N$ increases. Straight line guide for the eye. This plot implies that
    the final scaling function $f(y)$ is a simple quadratic function for large
    $N$ and $t \gg N$.}
  \label{fig:triangle} 
\end{figure}

In Fig.~\ref{fig:triangle} we plot the evolution of the empirical distribution
$\rho$ as a function of the number of bricklayers $N$ plotting as a function of
$\bar x^2$. For these detailed comparisons, we use up to $N=1024$ bricklayers, with
simulation times up to $t=32,384$ to ensure the convergence of the distribution
functions. 
We see that simulations with $N$ large converge to a triangular
function in this plot, indicating that the empirical distribution $\rho(x)$ is
parabolic at large times. 

From these curves, it is also possible to measure the width of the transition
region from large to small $\rho$. An empirical fit finds this width
$w\sim t^{2/3}/N^{0.4}$.  This is larger than the mean builder spacing
$t^{2/3}/N^{2/3}$. We have been unable to find an explanation for this law,
which goes beyond the continuum limit of the partial differential
equations.

\subsection*{Linearized equations}
\begin{figure}
  \includegraphics[width=.60\columnwidth]{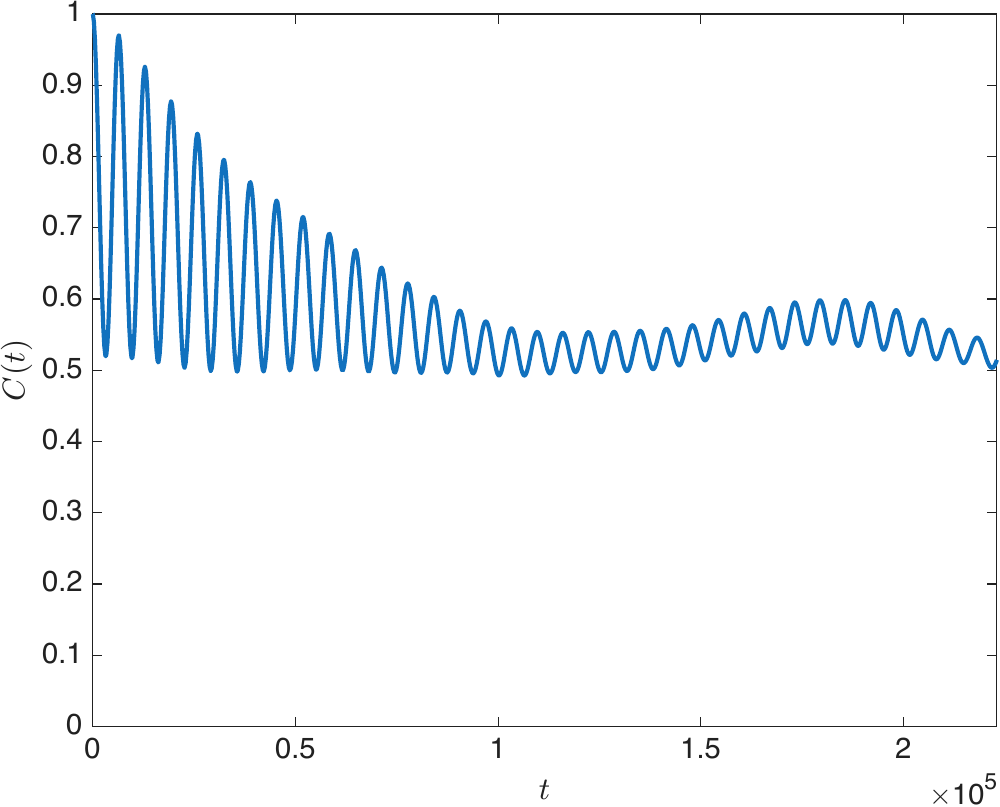}
  \caption{Simulation of a periodic system of $L=8192$ sites. $L/4$ bricklayers
    are placed in pairs, on random sites. The normalized autocorrelation,
    $C(t)$ of $P(t)$ 
    displays oscillations.}
  \label{fig:oscill}
\end{figure}

When there are many bricklayers distributed along the wall, and the
slope of the wall is small, we  write $\rho = \rho_0 + \tilde \varrho$ finding
the linearized equation for the density field,
\begin{equation}
  \frac {\partial^2 \tilde \varrho}{\partial t^2 } = \rho_0 \frac {\partial^2
    \tilde \varrho}{\partial x^2 }\, .
\end{equation}
This is a wave equation with propagation speed $c^2 = \rho_0$.
The bricklayers move in traveling waves when they are dense.  We perform
simulations by randomly placing pairs of builders (to impose even parity on
$(z_j+n_j)$) on $L/8$ sites of a flat wall of $L$ sites, imposing periodic
boundary conditions; we simulate sufficient time to come to a steady state, then
record the longest wavelength Fourier components of $\rho$.  We calculate
$P(t)= {c(t)}^2+{s(t)}^2$, with $c(t) = \sum_j n_j \cos{(q j)}$ and
$s(t) = \sum_j n_j \sin{(q j)}$ for $q=2 \pi/L$.  For a single traveling wave
$P(t)$ is constant. In the presence of two counter-rotating waves on a ring
there are moments of constructive and destructive interference.  In
Fig.~\ref{fig:oscill} we plot the autocorrelation of $P$ and observe coherent
oscillations. On longer time scales, the amplitude of the oscillations is
modulated, nonlinearities in the equations give rise to amplitude
exchange between modes

\subsection*{Conclusions}

We have found a (possibly non-unique) analytic solution to the partial
differential equation describing a generalized multi-walker true self-avoiding
motion. This solution is compatible with numerical simulations of the
model of~\cite{Toth2002}. Both the interface profile $h$ and the bricklayer density
$\rho$ are simple quadratic functions of the position, but are truncated at some
finite distance.

The true self-avoiding walk has been shown to have application in describing the
dynamics of non-reversible Monte Carlo simulations: Bricklayers are analogous to
``active particles'' in these methods, the wall height corresponds to the
displacement of atoms by the algorithm.  A multi-agent (parallel) version of
this algorithm has been described in~\cite{botao}. It will be interesting to see
if the ideas of the present paper can also be applied to such non-reversible
Monte Carlo algorithms in two or more dimensions with multiple agents. In many
implementations of non-reversible Monte Carlo, the active particle is resampled
at regular intervals in order to improve the ergodicity of the method, this
corresponds to additional source and sink terms in eq.~(\ref{eq:cont1})
\begin{equation}
  \frac{\partial \rho} {\partial t} + \frac{\partial (u  \rho)}{\partial
    x}=\lambda - \mu \rho \, ,
\end{equation}
where $\lambda$ is the creation rate of new builders, and $\mu$ a death rate.
We note also that with certain parameter values, non-reversible Monte Carlo
generates oscillating states, where the convergence of thermodynamic
averages becomes difficult to control. The oscillating states found in our
simulations of the bricklayer model are perhaps linked to this point.

\bibliography{avoid,scaling}
\end{document}